\documentclass[10pt,conference]{IEEEtran}

\usepackage{enumerate}

\usepackage{mathtools}
\usepackage{float}
\usepackage[center,singlelinecheck=false]{caption}
\usepackage{cite}
\usepackage{graphicx} %package to manage images
\usepackage{epstopdf}
\usepackage{amsmath} % assumes amsmath package installed
\usepackage{amssymb}  % assumes amsmath package installed
\usepackage{amsthm}
\usepackage[linesnumbered,ruled,vlined]{algorithm2e} % ,noend %remove "vlined" for puting "end"
\usepackage{nicematrix}

\usepackage[utf8]{inputenc}
\usepackage{color}
\usepackage{multirow}
\usepackage{optidef}
\theoremstyle{definition}
\newtheorem{example}{Example}

\newtheorem{lemma}{Lemma}

\newtheorem{definition}{Definition}

\newtheorem{remark}{Remark}

\DeclareMathOperator{\bin}{bin}
\DeclareMathOperator{\supp}{supp}

\DeclareMathOperator{\w}{w}

\newcommand{\wm}{\mathrm{w}_{\min}}

\newcommand{\Awm}{A_{\mathrm{w}_{\min}}}

\newcommand{\Aiwm}{A_{i,\mathrm{w}_{\min}}(\I)}

\newcommand{\Ki}{\mathcal{K}_i}

\newcommand{\B}{\mathcal{B}}

\newcommand{\I}{\mathcal{I}}

\renewcommand{\H}{\mathcal{H}}

\newcommand{\J}{\mathcal{J}}

\newcommand{\M}{\mathcal{M}}

\newcommand{\C}{\mathcal{C}}
\newcommand{\Ci}{\mathcal{C}_i}
\newcommand{\CI}{\mathcal{C}(\mathcal{I})}
\newcommand{\CiI}{\mathcal{C}_i(\mathcal{I})}

\newcommand{\bG}{\mathbf{G}}

\newcommand{\be}{\mathbf{e}}

\newcommand{\bc}{\mathbf{c}}

\newcommand{\bv}{\mathbf{v}}

\newcommand{\bzero}{\mathbf{0}}
\newcommand{\bone}{\mathbf{1}}

\newcommand{\bgi}{\mathbf{g}_i}
\newcommand{\bgj}{\mathbf{g}_j}
\newcommand{\bgm}{\mathbf{g}_m}

\newcommand{\bgh}{\mathbf{g}_h}
\newcommand{\bg}{\mathbf{g}}
\newcommand{\bp}{\mathbf{p}}
\newcommand{\bGN}{\boldsymbol{G}_N}
\newcommand{\bu}{\mathbf{u}}

\newcommand{\ft}{\mathbb{F}_2}

\usepackage{xcolor}
\def\BibTeX{{\rm B\kern-.05em{\sc i\kern-.025em b}\kern-.08em
    T\kern-.1667em\lower.7ex\hbox{E}\kern-.125emX}}

\usepackage{fancyhdr}

\IEEEoverridecommandlockouts                              % This command is only
\title{
Reverse PAC Codes: Look-ahead List Decoding
}
\author{
\IEEEauthorblockN{Xinyi Gu, Mohammad Rowshan, {\em Member, IEEE}, and Jinhong Yuan, {\em Fellow, IEEE}}

\IEEEauthorblockA{School of Electrical Eng. and Telecom., University of New South Wales (UNSW), Sydney, Australia\\ xinyi.gu@student.unsw.edu.au, \{m.rowshan,j.yuan\}@unsw.edu.au}
}

\begin{document}
\maketitle
\pagestyle{plain}
\thispagestyle{fancy}
\fancyhf{}
\lhead{\color{blue} 
%\footnotesize
To be presented at the 2024 IEEE International Symposium on Information Theory (ISIT 2024)\\
Session: TH4.R9: Topics in Modern Coding Theory 3
}
\cfoot{}

%%%%%%%%%%%%%%%%%%%%%%%%%%%%%%%%%%%%%%%%%%%%%%%%%%%%%%%%%%%%%%%%%%%%%%%%%%%%%%%%
\begin{abstract}

% Convolutional precoding in polarization-adjusted convolutional (PAC) codes can reduce the number of minimum weight codewords (a.k.a error coefficient) of polar codes. This can result in improving the error correction performance of (near) maximum likelihood (ML) decoders such as sequential decoders and sphere decoders. However, the convolutional precoding in PAC codes cannot be directly employed with sphere decoding. The reason is twofold: 1) the sphere decoding of polar codes is performed from the last bit - due to the lower rectangular shape of the polar transform. Whereas the shape of PAC codes generator matrix is no longer triangular. 2) To remove the former obstacle, one can modify the precoding step to get a lower-triangular shape. However, this may reduce the minimum distance of the resulting code due to the formation of unwanted cosets. %. %The reason is that the direction of decoding in sphere decoding is from the last bit to the first bit due to the lower rectangular shape of the polar transform. 
% In this work, we propose a selective convolutional precoding scheme with transposed precoding matrix to reduce the error coefficient while avoiding the reduction in the minimum distance.  %due to the formation of unwanted cosets. %Furthermore, we propose a method to construct a rate profile that further improves the error coefficient of reverse PAC codes. 
% The numerical results show a significant reduction in error coefficient, improving the block error rate by 0.2- 0.6 dB, depending on the code rate, in medium and high SNR regimes. 

Convolutional precoding in polarization-adjusted convolutional (PAC) codes is a recently introduced variant of polar codes. 
It has demonstrated an effective reduction in the number of minimum weight codewords (a.k.a error coefficient) of polar codes. This reduction has the potential to significantly improve the error correction performance. %of (near) maximum likelihood (ML) decoders. 
From a codeword formation perspective, this reduction has limitations. %in the PAC coding which depends on the rows of the generator matrix involved in the formation of codewords.  
%To overcome this limitation, 
Capitalizing on the understanding of the decomposition of minimum-weight codewords, this paper studies reverse precoding that can effectively reduce minimum-weight codewords more than in PAC codes. We propose a look-ahead list decoding for the reverse PAC codes, which has the same order of complexity as list decoding in PAC codes. 
%From the codeword formation point of view in cosets (as a result of row combination), one can observe that the reduction in the error coefficient in PAC coding is limited compared to polar codes and it depends on the distribution of frozen indices in the cosets of a code. This paper proposes a precoding scheme that defines \emph{frozen cosets} as a function of the input vector, resulting in the remarkable improvement of the error coefficient while avoiding the reduction in the minimum distance. 
Through numerical analysis, we demonstrate a notable reduction in error coefficients compared to PAC codes and polar codes, resulting in a remarkable improvement in the block error rate, in particular at high code rates. 
%This paper analyzes the formation of cosets in different pre-coding processes and proposes a different precoding scheme to significantly reduce the error coefficient while avoiding the reduction in the minimum distance. The numerical results show the expected improvement in the block error rate of short codes at low and high rates due to a significant reduction in the error coefficient with respect to PAC codes and polar codes. 
%The numerical results show the improvement of block error rate by 0.2-0.6 dB, depending on the code rate, in medium and high SNR regimes.
\end{abstract}

%%%%%%%%%%%%%%%%%%%%%%%%%%%%%%%%%%%%%%%%%%%%%%%%%%%%%%%%%%%%%%%%%%%%%%%%%%%%%%%%

\section{INTRODUCTION}
\label{sec:intro}
Polarization-adjusted convolutional (PAC) codes \cite{arikan2} are a variant of polar codes \cite{arikan} resulting from the convolutional pretransformation before polar coding. 
%Polar codes \cite{arikan} form a class of capacity achieving codes. However, they do not provide satisfactory error correction performance with a relatively low complexity successive cancellation (SC) decoding in finite block length. To address this drawback, SC list (SCL) decoding, introduced in \cite{tal}, provides a near maximum likelihood (ML) block error rate (BLER) at the cost of high computational complexity. Furthermore, it was recently proposed to concatenate convolutional codes and polar codes, resulting in a family of codes known as polarization-adjusted convolutional (PAC) codes \cite{arikan2}. 
The pre-transformation in PAC coding can reduce the number of minimum weight codewords of underlying polar codes due to the impact on the formation of minimum weight codewords \cite{rowshan2021precoding} and involvement of forzen coordinates carrying non-zero values. This reduction is expected to improve the performance of PAC codes under (near) ML decoders, such as the list decoder \cite{tal,rowshan-pac1,Yao_List_PAC_2020}, sequential decoders \cite{rowshan-pac1} and sphere decoder \cite{kahraman}, and could be considered for adaptation in 6G \cite{OJCOMS}. 
%
%In \cite{yuan_weight_2021} and \cite{rowshan-pac_enum}, the probabilistic weight distribution of pre-transformed polar codes and a fast deterministic method for the enumeration of minimum-weight PAC codes, respectively, were studied. 
The coset-wise study on the reduction of minimum weight codewords in PAC coding \cite{rowshan2023minimum} revealed that there are limitations to this reduction, particularly, when the contribution of frozen rows in codeword formation is lacking, or they do not possess a special property. One approach to overcome this drawback is to design a pre-coder that breaks these limitations. 
%
%Studies on precoding schemes have been limited. 
%
% In our previous work \cite{Xinyi_Selective}, convolutional precoding was applied in reverse order for PAC codes, resulting in a notable reduction in the number of minimum weight codewords for high-rate short codes. 
%
Different pre-transformations (or precoders) have been suggested in the literature such as the ones based on dynamic frozen bits or parity bits, the ones based on CRC bits, and finally, the ones based on the idea of pre-transformation in PAC coding \cite{ rowshan2021precoding, Chen_weighted_pac_2023, Liu_hamming_check_2023, wan_enhanced_weight_2023, Xinyi_Selective, Zunker_rowmerged_2023}. However, they do not perform the precoding from the perspective of the minimum-weight codeword formation. A detailed introduction of polar codes, PAC codes, and their variations can be found in \cite[Section VII]{OJCOMS}.
% In \cite{Samir_selectiveky_2020}, it was suggested to limit the precoding to frozen bits. %although it did not improve the error correction performance significantly. 
% A short-long bit-range precoder with two shift registers was proposed \cite{rowshan2021precoding}, where the first shift register was the conventional PAC precoder and the second was used to combine a subset of previous bits. Precodings of pac codes considering the weight distribution were studied in \cite{Chen_weighted_pac_2023, Liu_hamming_check_2023, wan_enhanced_weight_2023}.
%
In our previous work \cite{Xinyi2023improved},  %convolutional precoding was applied in reverse order for PAC codes, resulting in a notable reduction in the number of minimum weight codewords for high-rate short codes. 
%In this work, 
we introduced a precoding scheme for polar codes that can reduce the number of minimum weight codewords, considerably more than the PAC coding. The main idea is to involve more frozen rows in the formation of codewords through row combinations, particularly in cosets that are characterized as incapable of reducing the minimum weight codewords through precoding \cite[Lemma 2]{rowshan2023minimum}. %This scheme was initially suggested in \cite{Xinyi_Selective}.  

In this work, we further elaborate on the proposed precoding scheme in \cite{Xinyi2023improved} and name it reverse PAC (RPAC) codes. The shortcoming of \cite{Xinyi2023improved} was the lack of a low-complexity decoding algorithm for these codes. Since de-mapping of outer code (which is the inverse of precoding) is performed in the oppoite direction of decoding of inner code, the conventional successive cancellation list (SCL) decoding \cite{tal} cannot be used for the RPAC codes. To address this issue, we propose a look-ahead list (LA-SCL) decoding for RPAC codes which employs two decoding trees: 1) a tree similar to the list decoding of polar codes representing pre-transformed information, and 2) a tree representing the information sequence before the pre-transformation. The latter tree, called a look-ahead tree,  considers a few steps ahead of the current stage, taking into account the branch expansions corresponding to the reverse convolution. 
%an expanded one for vectors before precoding, referred to as the look-ahead tree, and another for vectors after precoding, similar to the conventional list decoders, referred to as the main tree. The look-ahead decoding tree explores the combinations generated by the reverse convolution, enabling the decoding of RPAC codes. 
Numerical results demonstrate that with the LA-SCL decoder, RPAC codes outperform CRC-Polar codes and PAC codes under conventional list decoding for high-rate short codes, while maintaining the same order of complexity. % to be larger at high rates and in high SNR regimes. %Needless to mention that this performance-improving scheme can be employed along with the complexity reduction techniques available in the literature though it is not in the scope of this work. %Hence, the contributions of this work can be summarized as follows:
% \begin{itemize}
%     \item The convolutional precoding in PAC codes are adapted to sphere decoding to reduce the error coefficient of polar codes. Moreover, the precoder is modified to avoid the challenging problem of reduction in the minimum distance which is not the case in other decoders, 
%     %\item For further reduction in the error coefficient, the rate profile is modified,
%     \item A detailed analysis of the impact of precoding on the error coefficient and the distribution of the minimum weight codewords in the cosets with respect to the choice of the convolutional polynomial is provided.
% \end{itemize}
%%%%%%%%%%%%%%%%%%%%%%%%%%%%%%%%%%%%%%%%%%%%%%%%%%%%%%%%%%%%%%%%%%%%%%
\section{PRELIMINARIES}\label{sec:prelim}
%\subsection{Notations}
%In this section, we first introduce some notations, and then we briefly review polar and PAC coding schemes, maximum likelihood decoding for a binary input additive white Gaussian noise (BI-AWGN) channel as well as sphere decoding algorithm as a candidate decoder for the proposed precoding scheme. 
%We denote by $\ft$ the finite field with two elements. The cardinality of a set is denoted by $|\cdot|$.  %The interval $[a,b]$ represents the set of all integer numbers in $\{x:a\leq x\leq b\}$. %The \emph{support} of a vector $\be = (e_1,\ldots,e_{n}) \in \ft^n$ is the set of indices where $\be$ has a nonzero coordinate, i.e. $\supp(\be) \triangleq \{i \in [1,n] \colon e_i \neq 0\}$ . %The indices of the vector elements may start from 0 or 1. 
Notations: %We denote the \emph{weight} of a vector $\be \in \ft^n$ by $w(\be)\triangleq |\supp(\be)|$. The support $\supp(\be)$ is the set of indices where $\be$ has a nonzero coordinate.
We denote the set of indices where vector $\be \in \ft^n$ has a nonzero coordinate by support $\supp(\be)$. The \emph{weight} of $\be$ is $\w(\be)\triangleq |\supp(\be)|$.
The all-one vector $\bone$ and the all-zero vector $\bzero$ are defined as vectors with all identical elements of 1 or 0, respectively. %The summation in $\ft$ is denoted by $\oplus$. Let $[\ell,u]$ denote the range $\{\ell,\ell+1,\ldots,u\}$ and bold letters denote vectors.
The (binary) representation of $i \in [0,2^n-1]$ in $\ft$ is defined as $\bin(i)=i_{n-1}...i_1i_0$, where $i_0$ is the least significant bit, that is $i = \sum_{a=0}^{n-1}i_a 2^a$. 
%A $K$-dimensional subspace $\cC$ of $\ft^N$ is called a linear $(N,K,d)$ \emph{code} over $\ft$ 
We use the operator $\backslash$ in $\mathcal{A}\backslash\mathcal{B}$ to subtract elements of the set $\mathcal{B}$ from $\mathcal{A}$.

\subsection{Polar Codes and PAC Codes} %%%%%%%%%%%%%%%%%%%%%%%%%%%%%%%%%%%%%%
\label{subsec:RMPolar}
Polar codes of length $N=2^n$ are constructed based on the $n$-th Kronecker power of binary Walsh-Hadamard matrix  
$\mathbf{G}_2 = 
{\footnotesize \begin{bmatrix}
1 & 0 \\
1 & 1
\end{bmatrix} }$, that is, $\bGN=\mathbf{G}_2^{\otimes n}=[\bg_0\;\;\bg_2\;;\cdots\;\;\bg_{N-1}]^T$ which we call it {\em polar transform} throughout this paper. %We denote the element $j$ in row $\bg_i$ of the polar transform by $g_{i,j}, i,j\in[0,N-1]$. 
A generator matrix of the polar code is formed by selecting the rows $\bg_i,i\in\I$ of $\bGN$. Then, $\CI$ denotes such a linear code. Note that $\I \subseteq [0,N-1]=[0,2^n-1]$. The characterization of the information set $\I$ for polar codes is based on the channel polarization theorem \cite{arikan} and the concept of \textit{bit-channel reliability}. %A polar code of length $N=2^n$ is constructed by selecting a set $\I$ of indices $i\in[0, N-1]$ with high reliability \cite{arikan}. 
The indices in $\I$ are dedicated to information bits, while the indices in $\mathcal{I}^c \triangleq [0, N-1]\setminus \I$ are used to transmit a known value, '0' by default, which are called \emph{frozen bits} and the corresponding rows are frozen rows. 
%Regardless of the method, we use for forming the set $\mathcal{I}$ for a polar code, the bit-channels with indices in the set $\mathcal{I}$ must be more reliable than any bit-channels in $\mathcal{I}^c$. %The notation $W^{(j)}_N\preceq W^{(i)}_N$ is used to say that the bit-channel $i$ is more reliable than bit-channel $j$. 

In PAC coding \cite{arikan2}, however, we have a pre-transformation stage before polar coding where the input vector $\mathbf{u}=[u_0,\ldots,u_{N-1}]$ for polar coding is obtained 
%To improve the distance properties of polar codes, it was suggested in \cite{arikan2} to obtain the input vector $\mathbf{u}=[u_0,\ldots,u_{N-1}]$
%\subsection{PAC Codes}\label{ssec:PAC} %%%%%%%%%%%%%%%%%%%%%%%%%%%%%
%The recently introduced polarization-adjusted convolutional (PAC) coding scheme \cite{arikan2} can reduce the number of min-weight codewords. %This reduction is a result of inclusion of rows in $\I^c$ in the generation of codewords in the cosets \cite{rowshan2021precoding}. Note that the convolutional precoding does not change set $\mathcal{B}$. In this section, we study how precoding further reduces the number of min-weight codewords.
by a convolutional transformation using the binary generator polynomial of degree $s$, with coefficients $\mathbf{p}=[p_0,\ldots,p_s]$ as follows: 
\begin{equation}\label{eq:precoding}
    u_i = \sum_{\ell=0}^s p_\ell v_{i-\ell},
\end{equation} 
where $\mathbf{v}=[v_0,\ldots,v_{N-1}]$ is the vector constructed based on $\I$. 
%This convolutional transformation combines $s$ previous input bits with the current input bit $v_i$ to calculate $u_i$.
This coding scheme is called {\em polarization-adjusted convolutional (PAC)} coding. %The parameter $s$ is known as the {\em memory} of the shift register and by including the current input bit  we have the {\em constraint length} $s+1$ of the convolutional code. 
The convolution operation can be represented in the form of an upper triangular matrix \cite{rowshan2021precoding} where the rows of the {\em pre-transformation matrix} $\mathbf{P}$ are formed by shifting the vector $\mathbf{p} = (p_0,p_1,\ldots p_s)$ one element at a row. 
Note that $p_0=p_s=1$ by convention. Then, we can obtain $\mathbf{u}$ by matrix multiplication as $\mathbf{u}=\mathbf{v}\mathbf{P}$.
% \begin{equation}
% \label{eq:conv_gen}
% \mathbf{P}=\begin{bNiceMatrix}
% p_0 & p_1 & \Cdots & p_m & 0 & \Cdots & & 0 \\
% 0 & p_0 & p_1 & \Cdots & p_m & & & \\
% \Vdots & \Ddots & \Ddots & \Ddots & & \Ddots & \Ddots &\Vdots \\
%  & & & & & & & 0 \\
%  & & & & p_0 & p_1 & \Cdots & p_m \\
% \Vdots & & & & \Ddots & \Ddots & \Ddots &\Vdots \\
%  & & & & & & p_0 & p_1 \\
% 0 & \Cdots & & & & & 0 & p_0
% \end{bNiceMatrix}
% \end{equation}

Due to this precoding, we would have $u_i\in\{0,1\}$ for $i\in\mathcal{I}^c$, indicating that $u_i$ correpsoding to a frozen $v_i=0,i\in\I^c$ may no longer be fixed. 
Consequently, the vector $\mathbf{u}$ is mapped to codeword $\mathbf{x}=\mathbf{u}\mathbf{G}_N$. Overall, we obtain $\mathbf{x}=\mathbf{v}\mathbf{P}\mathbf{G}_N$. %The concept of dynamic frozen bits for polar codes was initially introduced in \cite{Trifonov_2013_Dynamic}.
It was analytically shown in \cite{rowshan2021precoding,rowshan2023formtion} that by convolutional pre-transformation, the number of minimum weight codewords, a.k.a error coefficient which is denoted by $\Awm$ where $\wm$ is the minimum weight, may significantly decrease relative to polar codes (without pre-transformation). Hence, from the union bound \cite[Sect. 10.1]{lin_costello}, we expect that this reduction potentially improves the block error rate (BLER) of a (near) maximum likelihood decoding for a binary input additive white Gaussian noise (BI-AWGN) channel, particularly at high signal-to-noise ratios (SNRs). 
\subsection{Minimum Weight Codewords in Cosets}%the Proposed precoding} %%%%%%%%%%%%%%%%%%%%%%%%%%%%%%%%%%%%
%\mohammad{We should remove the frozen coset term. }
In the conventional PAC coding, forward convolution as per \eqref{eq:precoding} is performed. Although forward convolution can reduce the number of codewords of minimum weight relative to polar codes \cite{rowshan2021precoding}, it has its own limitations. To show the limitations, we first partition all codewords, excluding the all-zero codeword, of a polar code $\C(\I)$ into \emph{cosets} defined as: %follows \cite[Defenition 3]{rowshan2023formtion}:
\begin{definition} \label{def:coset}
    %(\cite{rowshan-pac1}) 
    Cosets: Given information set $\I \subseteq [0,N-1]$ for a polar code, we define the set of codewords $\CiI\subseteq \CI$ for each $i\in \I$ in a coset of the subcode $\C(\I \setminus [0,i])$ of $\C(\I)$~as 
    \begin{equation}\label{eq:Ci}
        \CiI \triangleq \left\{\bgi+\sum_{h\in \H} \bgh \colon \H \subseteq \I \setminus [0,i]\right\}\subseteq \C(\I),
    \end{equation}
    where $\bgi$ is the \emph{coset leader}. We denote the number of minimum weight codewords of the coset $\Ci$ by $\Aiwm$. The total number of minimum weight codewords for a polar code $\C(\I)$ is $\Awm=\sum_{i\in\I}\Aiwm$.
\end{definition}

Observe that the coordinate of the first non-zero element in vector $\bu$, $i=\min\{\supp(\bu)\}$, while encoding by $\mathbf{x}=\mathbf{u}\mathbf{G}_N$ plays a key role in classifying the codewords into cosets. According to \cite[Lemma 1]{rowshan2023minimum}, this coordinate remains the same after precoding by $\bu=\bv\mathbf{P}$. That is, 
\begin{equation}\label{eq:preservation}
    \min\{\supp(\bu)\}=\min\{\supp(\bv)\}.
\end{equation}
Nevertheless, the resulting $u_j$ for $j>i,j\in\I^c$ might be $u_j\not=0$, unlike in polar coding. %Hence, although due to retaining $i=\min\{\supp(\bu)\}$ in PAC coding, the minimum distance remains the same as polar codes, that is,
% \begin{equation}
%     d_{min} = w_{min} = \min(\{w(\bgi):i\in\I\}),
% \end{equation}
This difference may impact the number of minimum weight codewords in the cosets due to the inclusion of rows $\bg_j$ for $j\in\I^c\cap[i,N-1]$ in row combinations. Note that the weight of the codewords in a coset $\Ci$ depends on the set of row indices $\H=\{h:h>i\}$ in \eqref{eq:Ci}. Furthermore, according to \cite[Corollary 5]{rowshan2023formtion}
\begin{equation}\label{eq:geq_wi}
    \w(\bgi+\sum_{h\in\mathcal{H}}\mathbf{g}_h)\geq \w(\bgi),
\end{equation}
there exists minimum weight codeword $\bc\in\Ci$, where $\w(\bc)=\wm$, only if $i\in\mathcal{B}$ where set $\mathcal{B}$ is defined as
\begin{equation}
    \mathcal{B}=\{j:j\in\I,w(\bg_j)=\wm\}.
\end{equation} 
Observe that we have $\H\subseteq[i+1,N-1]$ in PAC coding whereas in polar coding, we have $\H\subseteq[i+1,N-1]\backslash\I^c$. %Removing the constraint on $\H$ and considering all rows may give different codewords.  
The minimum distance of the PAC codes is \cite[Lemma 1]{rowshan2023minimum} 
\begin{equation}
    d_{\min} = \wm = \min(\{\w(\bgi):i\in\I\}).
\end{equation}

% Now, as the minimum distance of a polar code equals the minimum weight of the non-frozen rows in the polar transform, that is, 
% \begin{equation}
%     d_{min} = w_{min} = \min(\{w(\bgi):i\in\I\}).
% \end{equation}

% On the other hand, according to Corollary 5 in \cite{rowshan2023formtion}, we~have
% \begin{equation}\label{eq:geq_wi}
%     w(\bgi\oplus\bigoplus_{j\in\mathcal{H}}\mathbf{g}_h)\geq w(\bgi),
% \end{equation}
% where $\mathcal{H}\subseteq [i+1,2^{n}-1]$. Hence, the number of the minimum weight codewords denoted by $\Aiwm$, can be formed in the cosets $\CiI$ where the coset leader has weight $w(\bgi)=w_{min}$.

% In the forward convolution used in the conventional PAC codes, since for every $\CiI$, we have $i\in\I$, the minimum distance of polar codes is preserved. 

% Observe that in polar coding, we always have $b=0$ at coordinates $i\in\cI^c$ whereas $b\in\{0,1\}$ in PAC coding depending on the choice of coefficient vector $\bp$. 
%\textcolor{blue}{
According to \cite[Theorem 1]{rowshan2023formtion}, the minimum weight codewords are uniquely formed by the following row combinations:
\begin{equation}\label{eq:wt_gi_gj_gm}
    \w\big(\bgi+\sum_{j\in\J}\bgj +\sum_{m\in\M(\J)}\bgm\big) =  \wm,
\end{equation}
% \begin{equation}\label{eq:wt_gi_gj_gm}
%     w\big(\bgi\oplus\bigoplus_{j\in\J}\bgj \oplus \bigoplus_{m\in\M(\J)}\bgm\big) =  w_{min},
% \end{equation} 
%where $\bgi$ is the leading row, $\bg_j, j\in\J$ are the core rows, and $\bg_m, m\in\M(\J)$ are the balancing rows. 
where $\w(\bg_i)=\wm$, $\J\subseteq\Ki$ and $\Ki$ is \cite[Lemma 2.a]{rowshan2023formtion}
\begin{equation}\label{eq:set_J}
\Ki \triangleq \{j \in \I\backslash[0,i]\colon |\supp(j)\backslash\supp(i)|=1\}.
\end{equation}
As a result, every subset of $\Ki$ along with other rows in \eqref{eq:wt_gi_gj_gm} form a minimum weight codeword. 
The number of subsets of $\Ki$ is given by $2^{|\Ki|}$. Given $\B\triangleq\{i\in\I:\w(\bgi)=\wm\}$, the total number of minimum-weight codewords of the polar code will be $\sum_{i\in\B}2^{|\Ki|}$. 
%
%Since we can have $2^{|\Ki|}$ subsets of any set in total, we can have this many minimum-weight codewords. 
The set $\M(\J)$ is a function of the set $\J$ and every $m\in \M(\J)$ has the property (see \cite[(9),(10)]{rowshan2023formtion} for a detailed definition of $\M(\J)$):
\begin{equation}\label{eq:set_M}
\M(\J)\!\subseteq\!\{m\!>\!i:|\supp(\bin(m))\backslash\supp(\bin(i))|\!>\!1\}.
\end{equation} 
%
% The number of minimum weight codewords that are generated by the leading row $\bgi$ is denoted by $A_{i, \wm}$.
% where the general properties of sets $\J$ and $\M$ are 
% \begin{equation}\label{eq:set_J}
% \J\!\subseteq\!\{j\!>\!i\!:\!|\supp(\bin(j))\backslash\supp(\bin(i))|\!=\!1\},
% \end{equation} 
% According to \cite{rowshan2023formtion} and \cite{rowshan-pac_enum}, the minimum weight codewords are uniquely formed by row combinations in matrix $\bG_N$ as
% \begin{equation}\label{eq:wt_gi_gj_gm}
%     w\big(\bgi\oplus\bigoplus_{j\in\J}\bgj \oplus \bigoplus_{m\in\M(\J)}\bgm\big) =  w_{min},
% \end{equation} 
% where the general properties of sets $\J$ and $\M$ are 
% \begin{equation}\label{eq:set_J}
% \J\!\subseteq\!\{j\!>\!i\!:\!|\supp(\bin(j))\backslash\supp(\bin(i))|\!=\!1\},
% \end{equation} 
% \begin{equation}\label{eq:set_M}
% \M\!\subseteq\!\{m\!>\!i:|\supp(\bin(m))\backslash\supp(\bin(i))|\!>\!1\}.
% \end{equation} 
%Note that for every set $\J$, there exists a unique set $\M$ with the general property in \eqref{eq:set_M}. In this paper, we do not need to know how set $\M$  is formed as a function of $\J$ and the difference between these two sets suffices for our discussions. %Note that the elements of set $\J$ and $\M$ are in $(i,N-1]$.
The relation \eqref{eq:wt_gi_gj_gm} can be extened such that the sets $\J$ and $\M$ also intersect with $\I^c$ (see \cite[(18)]{rowshan2023minimum}). This is useful when considering the impact of precoding.
% \begin{remark}\label{rem:w_min_sabotage}
%     According \eqref{eq:wt_gi_gj_gm},  observe that if for some $\J$, we do not have the corresponding set $\M(\J)$  (although we might have $\M\not=\emptyset$), the generation of minimum weight codewords in coset $\Ci$ can be avoided. Conversely, this is the case if, for some $\M$, there is no corresponding $\J$ (including the case $\J=\emptyset$). This frequently happens as a result of precoding. %where the control over the value of $u_i,i\in\I^c$ is limited in the mapping $v_i\rightarrow u_i$ knowing $v_i=0$.
% \end{remark}
%}
Now, let us see the main limitation of the forward convolution in PAC coding, %According to \cite[Remark 2]{rowshan-pac_enum}, there are two categories of cosets %characterized based on any $j\in\I^c$ for $j>i$ 
that forward convolution cannot reduce their minimum weight codewords.  
\begin{lemma}\label{lem:incapble}
    (\cite[Lemma 2]{rowshan2023minimum}) For any coset $\CiI$ where
    \begin{enumerate}
        \item $\I^c \cap(i, N-1]=\emptyset$, or
        \item $|\supp(\bin(f)) \!\backslash\! \supp(\bin(i))|\!=\!1, \forall f\!\in\!\left(\I^c\!\cap\!(i,N\!-\!1]\right)$,
    \end{enumerate}
    we have
    $$
    A_{i, \wm}(\bG, \I)=A_{i, \wm}(\mathbf{P G}, \I) .
    $$
    
    In other words, any cosets $\mathcal{C}_i$ where there is no frozen row $\mathbf{g}_f$ for $f \in \mathcal{I}^c \cap(i, N-1]$ such that $|\supp(\bin(f)) \backslash \supp(\bin(i))|>1$, we get $A_{i, \wm}=$ $2^{\left|\mathcal{K}_i\right|}$ in the PAC coding, independently of the choice of $\mathbf{p}$.
    \end{lemma}
%The obvious category is where we have $\H\cap\I^c=\emptyset$ for a coset, i.e., there exists no row $\bg_f$ for $f\in\I^c$ in the coset. The second category is where $\H\cap\I^c\not=\emptyset$ but for every $f\in\H\cap\I^c$, we have $|\supp(\bin(f))\backslash\supp(\bin(i))|=1$. 
%Hence, there is a lower bound for the error coefficient in the conventional PAC coding with forward convolution. 

%In the next section, we propose a different precoder that overcomes the limitations of the forward convolution.
% \begin{remark}
%     As forward precoding regardless of the choice of polynomial $\bp$ is not effective on some cosets characterized in \cite[Remark 2]{rowshan-pac_enum}, the capability of reducing the error coefficient of polar codes by forward precoding is limited.
% \end{remark}

\section{Reverse PAC (RPAC) Codes}\label{ssec:gen_approach} %%%%%%%%%%%%%%%%%%%%%%%%%%%%%%%%%%%%%%%%%%%%%%%
%As discussed the first non-zero element in $\bu$ defines the coset leader and the coset leader indicates whether the coset includes any minimum weight codewords or not, as per \eqref{eq:geq_wi}. 
To tackle the limitations described for the characterised cosets in Lemma \ref{lem:incapble} and further reduce the total number of minimum weight codewords $A_{\wm}$, the general idea is to change the coset leader $\bgi$ in \eqref{eq:Ci} through a different precoding scheme such that the coordinate of the first non-frozen element in $\bu$ is not preserved, unlike \eqref{eq:preservation}. That is, we need a pre-transformation matrix $\mathbf{P}'$ that gives $\bu=\bv\mathbf{P}'$ where 
\[
    \min\{\supp(\bu)\}\not=\min\{\supp(\bv)\}.
\]
%The matrix $\mathbf{P}'$ is required to be designed such that 
To further refine the problem, for any pair of $(i,i')$ where $i'=\min\{\supp(\bv)\}$ and $i=\min\{\supp(\bu)\}$, we need to have 
\begin{enumerate}
    \item $w(\bg_i)\!\geq\!w(\bg_{i'})$. Observe that if we have $w(\bg_i)\!<\!w(\bg_{i'})$, according to \eqref{eq:geq_wi}, the minimum distance $d_{min}$ after precoding might decrease. Note that the resulting coset may be led by a frozen row $\bg_i, i\in\I^c$.
    \item $i < i'$. Observe that if we have $i > i'$, we may lose some of the information bits.
\end{enumerate}
% \begin{proposition}
% Given a coset $\Ci$ for $\i\in\mathcal{B}$
% \end{proposition}
%\textcolor{blue}
{\begin{remark}
      Letting $\min\{\supp(\bu)\}<\min\{\supp(\bv)\}$ moves the coset leader from $\bg_{i'}$ to $\bg_{i}$ where $i<i'$. This will affect the minimum weight codewords in the coset $\Ci$  as follows:
    \begin{enumerate}[a)]
        \item If $\w(\bg_i) > \w(\bg_{i'})$, according to \eqref{eq:geq_wi}, we have $A_{i,\wm}=0$. 
        \item If $\w(\bg_i) = \w(\bg_{i'})$ and the conditions of Lemma \ref{lem:incapble} are violated, then $A_{i, \wm}(\mathbf{P G}, \I)<A_{i, \wm}(\bG, \I)$.
        %there may exist some $\bg_f$ where $f\in\I^c\cap[i,N-1]$ and $|\supp(\bin(f))\backslash\supp(\bin(i))|>1$ (see \cite[Remark 2]{rowshan-pac_enum}). 
        %the error coefficient $\AiwmNoI$ of coset $\C_{i}$ can reduce, relative to coset $\C_{i'}$ where there exists no such $\bg_f$.
    \end{enumerate}
\end{remark}
%}
%Now, let us rewrite  \ref{def:coset} to be consistent with the proposed way of forming the cosets.

\subsection{Practical Approach for Designing the Pre-transform}
There might be different ways to implement the general approach proposed above. In this section, we propose a practical scheme to realize the proposed approach. 

Our scheme is based on designing a mapping function $\bv\rightarrow\bu$ such that we get $\min\{\supp(\bu)\}<\min\{\supp(\bv)\}$, and then form the corresponding pre-transformation  matrix $\mathbf{P}'$.  The requirements for the mapping are as follows:
\begin{itemize}
    \item $u_i=1$ for $i<i'$ where $i'=\min\{\supp(\bv)\}$,
    \item $u_j\in\{0,1\}$ for $j>i$,
    \item $u_j \leftarrow v_j+\sum_{k\in\mathcal{K}} p_k v_k$ where $\mathcal{K}\subset[i',N-1]\backslash \{j\}$ and $p_k\in\{0,1\}$.
\end{itemize}
To make the scheme tractable, we constrain the scope of $\mathcal{K}\subset[i',N-1]\backslash \{j\}$ by selecting $\mathcal{K}=[j+1,j+s]$ for some $s$. Hence, we can write this mapping function in $\ft$ as 
\begin{equation}%\label{eq:precoding_rev}
    u_i = \sum_{\ell=0}^s p_\ell v_{i+\ell}.
    \label{eq:FC_PAC}
\end{equation}
The vector $\bp=[p_0,p_1,\cdots,p_{s}]$ is similar to the one used in the conventional precoding of PAC codes. However, the matrix form of the proposed precoder has a lower triangular shape; that is, $\mathbf{P}'=\mathbf{P}^T$. %follows: 
% \begin{equation}
% \label{eq:conv_gen_rev}
% \mathbf{P}' =
% \begin{bNiceMatrix}
% p_0 & 0 &  & & & & \Cdots & 0 \\
% p_1 & p_0 & \Ddots & & & & & \\
% \Vdots & \Ddots & \Ddots & & & & &\Vdots \\
% p_m & \Cdots & p_1 & p_0 & & & & \\
% 0 & & & & & & & \\
% \Vdots & \Ddots & \Ddots & & \Ddots & \Ddots & \Ddots &\Vdots \\
%  & & & p_m & \Cdots & p_1 & p_0 & 0 \\
% 0 & & \Cdots & 0 & p_m & \Cdots & p_1 & p_0
% \end{bNiceMatrix}
% \end{equation}
From linear algebra, we know that the product of two lower (or upper) triangular matrices is a lower (or upper) triangular matrix. Hence, knowing $\mathbf{G}_N$ is a lower triangular matrix, we can conclude that $\mathbf{P}'\mathbf{G}_N$ is also a lower triangular matrix. 

% Now, Let us revisit the coset defined in Definition \ref{def:coset}. Here we rewrite it as follows to be consistent with the proposed scheme:
% \begin{equation}
%     \Ci \triangleq \left\{\bg_i\oplus\bigoplus_{h\in \H} \bgh : \H \subseteq [i\!+\!1,N\!-\!1]\right\}.
% \end{equation}
%\mohammad{What was your idea to further improve Reverse PAC codes? You apply it now.}
% {\color{blue}
% \begin{lemma}\label{lem:dmin_decrease}
    
%     $d_{\min}(\mathbf{P}^{\prime}\bG,\I)\leq d_{\min}(\mathbf{P}\bG,\I)$.
% \end{lemma}

% \begin{proof}
%     positions of i

%     cases:

%     depends on $i'-s$
    
%     $\w(\mathbf{g}_{i'-s}) \ge \dm$

%     $\w(\mathbf{g}_{i'-s}) < \dm$
% \end{proof}
% }

Furthermore, to fulfill the first requirement mentioned in Section \ref{ssec:gen_approach} as $\w(\bg_i) \ge \w(\bg_{i'})$ and to avoid the reduction in $\wm$, %as discussed in the previous section 
we need to constrain the precoder as follows:
\begin{equation}
\text{$u_i$}=
    \begin{dcases*}
        \sum_{\ell=0}^s p_{\ell} v_{i+\ell} & if $\w(g_i) \ge \wm$\\
        v_i & otherwise \\
    \end{dcases*}.
    \label{eq:RPAC_conv}
\end{equation}

As can be seen in \eqref{eq:RPAC_conv}, the mapping to $u_i$ is based on $[v_i\;v_{i+1}\;\cdots\;v_{i+s}]$ which are placed ahead of coordinate $i$. Thus, the convolution is performed in \emph{reverse} direction, the opposite of the forward direction in the PAC coding. Hence, this scheme is called reverse PAC or in short RPAC. We use the notion RPAC($s+1$) to denote the constraint length, $s+1$, of the reverse precoder. 
Now, let us see a simple example that illustrates the difference between precoding in PAC codes and the proposed RPAC codes.

\begin{example}\label{fig:example_rev}
    Consider the polar code (64,14) with $d_{min}=16$ and $\I=\{31,46,47,51,53\!-\!55,57\!-\!63\}$ where '$-$' indicates a range of integers. Given a vector $\bv$ with $v_i=0$ for every $i\in\I$ except for $v_{54}=1$, then the minimum weight of codewords %are the codewords in terms of row combinations, 
    obtained by row combinations for
    polar coding where $\bp=[1]$, PAC coding with $\bp=[1\;1\;0\;1\;1\;0\;1\;1\;0\;1]$, and RPAC coding with the same $\bp$ are: 
    \begin{equation*}
        \text{Polar:}  \w(\bv\mathbf{P}\bG_N=\bg_{54})=16,\vspace{-5pt}
    \end{equation*}
    \begin{equation*}
        \text{PAC:}  \w(\bv\mathbf{P}\bG_N=\bg_{54}+\bg_{56})=16,\vspace{-5pt}
    \end{equation*}
    \begin{equation*}
        \text{RPAC:}  \w(\bv\mathbf{P}'\bG_N=\bg_{43}+\bg_{54})=24.
    \end{equation*}    
    Note that in the RPAC coding, the frozen row $\bg_{43}$ with the index $43<54$ is involved in the row combination. This effectively increases the weight of the RPAC codeword to 24. % While the codeword weight remains 16 in polar and PAC coding, RPAC codeword has a weight of 24. 
\end{example}
\section{Look-ahead List Decoding for RPAC} 
Reverse PAC codes have an advantage over PAC codes in terms of the number of minimum weight codewords. However, this advantage brings about challenges in decoding due to the opposite directions of pre-transformation and decoding. 
%As we observed in the previous section, the pre-transformation is carried out from the last bit, 
%while existing SC-based decoding procedures are essentially started from the first bit and move forward. 
A trivial solution is to employ generic decoders to guess the coordinate(s) of errors. Alternatively, sphere decoding can be considered, which is performed from the last bit towards the first bit. However, these candidate decoders exhibit high and variable computational complexity, and consequently high and non-constant latency and energy consumption. 
Here, we propose a look-ahead list decoding algorithm that expands the $\bv$ sequence %$\hat{v}_0^i$ 
on each path by $s$ bits ahead of the current stage $i$. %, hence called look-ahead list decoding. 
We use $L$ to denote the list size of the LA-SCL decoder. 
%\xinyi{expands the $v_0^i$-sequences on each path by $s$ bits ahead of the current stage $i$}

The look-ahead list decoding can be considered as the expansion of search trees. We use two trees to represent the expansions: 1) a main tree for $\bu$, corresponding to the tree in conventional list decoding, where the leaves represent stage $i$. 2) a parallel binary tree for $\bv$, affiliated with the main tree, looks ahead for stage $i+s$. We refer to this tree as the \emph{look-ahead tree}. This tree maintains the same number of paths as the main tree but extends $\bv$-values for $s$ stages ahead, corresponding to each path. 
It's important to note that the look-ahead tree only represents binary values of $\bv$, while the $L$ independent SC processes with different intermediate LLRs are retained on the main tree.
During decoding, at stage $i$ of the main tree, the binary tree for $\bv$ expands first by examining stage $i+s$. If $i+s \in \I$, each path expands with two possibilities: $\hat{v}_{i+s} = 0$ and $\hat{v}_{i+s} = 1$. Considering the sequences of: 
\begin{equation}\label{eq:v_seq}
[\hat{v}_{i}\;\; \hat{v}_{i+1}\;\; \cdots\;\; \hat{v}_{i+s}],
\end{equation}
the estimated $\bv$ becomes the input to the convolutional transform in \eqref{eq:RPAC_conv}. The output $\hat{u}_i$ corresponds to the paths on the main tree. 
The path metrics are computed based on the values of $\hat{u}_i$, which are then fed back to the SC process to conduct the partial sums for intermediate LLRs. 
As we move from stage $i$ to stage $i+1$ on the main tree, we extend the leaves of the look-ahead part of the tree by one branch from $i+s$ to $i+s+1$ and append the stage $i+1$ to the main tree. The convolution in \eqref{eq:RPAC_conv} for $\hat{u}_{i+1}$ and partial sums for intermediate LLRs are computed accordingly. 

%Path metrics for each path are retained on the main tree. 
% Path metrics are computed based on the values of $\hat{u}_i$ and fed back to the SC process to compute the partial sums. The intermediate LLRs for each path are also retained on the main tree. 
%When the associated tree reaches the last stage, i.e. $i+s = N$, both of the trees stop expanding.

%As a result, 
%A sequence $\mathbf{v}$, the input to the convolutional transform in \eqref{eq:RPAC_conv}, corresponds to a path on an irregular binary tree.  

%%The fundamental idea behind the look-ahead list decoding is as follows. During decoding (that is, the expansion of the tree), we use two trees to represent the expansions for $\bu$ and $\bv$:the tree will have two parts:
%1) the main tree with origin at step $i=0$ and the leaves representing stage $i$. This part is similar to the one in conventional list decoding, where we compute path metrics based on the values of $\hat{u}_i$ and feedback this value to the SC process to compute the partial sums. In addition, we keep the intermediate LLRS for each path on the main tree. 2) Look-ahead paths: Each path on the main tree will have an extension of $s$ stages, which looks ahead and considers the sequences of 
% \begin{equation}\label{eq:v_seq}
%     [\hat{v}_{i+1}\;\; \hat{v}_{i+2}\;\; \cdots\;\; \hat{v}_{i+s}],
%  \end{equation}
%satisfying the constraint in \eqref{eq:RPAC_conv}. 
%This binary pruned subtree belongs to the data structure of every path.

For the LA-SCL decoder, at stage $i=0$, the affiliated tree needs to look ahead for $s$ stages, requiring initialization. 
The initialization of the trees is determined by the number of information bit coordinates within $v_0^s$, denoted by $\nu$. This initialization has two conditions: a) If the first information bit coordinate, denoted by $\I_0$, satisfies $\I_0 \ge s$, the look-ahead tree undergoes an expansion process similar to the conventional expansion of the list decoder for PAC codes. When the looking-ahead sequence $\hat{v}_{i+s}$ encounters $\I_0$, the look-ahead tree initiates expansion into two paths of $\hat{v}_{i+s}=0$ and $\hat{v}_{i+s}=1$. 
b) However, if $\I_0 < s$ in very high code rate conditions, the look-ahead tree needs to consider all the possible combinations of $[\hat{v}_{\I_0}\; \cdots\; \hat{v}_{s}]$. It will be initialized into a tree with a depth of $s-\I_0+1$, and the minimum required list size becomes $L_{min} = 2^{\nu}$. 
% The main tree $u_i$ can be expanded based on the expansion of the look-ahead subtree accordingly, so as intermediate LLRs.

% \textcolor{blue}{the explanation on how the decoding end is missing}

Note that the look-ahead list decoding does not increase the list size compared to the conventional list decoding. 
Every path on the main tree for $\hat{u}_i$ corresponds to the paths on the look-ahead tree for $\hat{v}_{i+s}$ in a one-to-one manner. The SC process is performed based on the main tree and the intermediate and decision LLRs are calculated for the $L$ paths up to stage $i$. 
In stage $i+1$ (corresponding to the main tree), we extend every path of the look-ahead tree by appending $\hat{v}_{i+s+1} \in\{0,1\}$. Then, the corresponding $\hat{u}_{i+1}$ on the main tree are computed based on the expanded $\hat{v}_{i+s+1}$. The number of the paths on both trees becomes $2L$. When $L\ge L_{max}$, paths with the largest path metrics on the main tree, along with the corresponding paths on the look-ahead tree, are pruned. 
Therefore, since the tree expansion of the main tree is similar to conventional list decoding and the look-ahead tree is expanded in parallel to keep track of $\bv$-values ahead of stage $i$, the order of complexity remains the same, i.e. $O(L\cdot N \log_2 N)$.

To clarify this matter, let's consider an example.% depicted in Fig. \ref{fig:RPAC_LASCL_example}.
\begin{example}\label{ex:v_seq}
    %Suppose that we are in the stage $i \in \I$ where $\w(\bg_i)\geq \wm$ and performing look-ahead list decoding with list size $L=4$. Given $s=3$ and $\bp=[1\;0\;\;1\;\;1]$, we obtain $u_i$ using the constraint in \eqref{eq:RPAC_conv} for $\times2^{s}$ possible sequences of \eqref{eq:v_seq} for $L$ look-ahead subtrees. Then, based on $u_i$ we compute the path metrics of 
    % \begin{equation*}
    %     \begin{multlined}
    %         [0\;\; 0\;\; 0],\;\;\; [0\;\; 1\;\; 1],\;\;\;\;\;
    %         [1\;\; 0\;\; 0],\;\;\; [1\;\; 1\;\; 1].
    %     \end{multlined}
    % \end{equation*}

    Given $s=3$ and $\bp=[1\;0\;\;1\;\;1]$, suppose that we are in stage $i = 0$ and performing look-ahead list decoding with list size $L=4$. We consider two initialization conditions with different information sets: $\I = \{3,\; 4\}$ and $\I = \{2,\;3,\; 4\}$. The corresponding decoding trees are illustrated in Fig. \ref{fig:RPAC_LASCL_example}.
    
    For $\I = \{3,\; 4\}$, look-ahead tree initiates expansion until it reaches $\hat{v}_3$. An example of normal initializations of the trees are depicted in Fig. \ref{fig:RPAC_LASCL_example} (a). At stage $i = 0$, the initialization of the look-ahead tree is marked in green. 
    While for $\I = \{2,\; 3,\; 4\}$, two information bit coordinates are included within the range of $\I_0$ to $s$, i.e. $\nu = 2$. Consequently,
    all possible paths of $[\hat{v}_2,\; \hat{v}_3]$ needs to be listed. This special initialization, where $\I_0 < s$, is shown in Fig. \ref{fig:RPAC_LASCL_example} (b).
    The main tree for $\hat{u}_i$ is expanded based on the look-ahead trees, resulting in the same number of paths. Moving forward to stage $i = 1$, $\hat{v}_{4}$ is expanded into two possible values of $\hat{v}_{4} \in \{0,\; 1\}$ for all existing paths, which is the same as the conventional SCL decoder, but with $s$ bits looking ahead. The expansion for stage $i = 1$ is shown in yellow in the figure. For the example in Fig. \ref{fig:RPAC_LASCL_example} (b), at level $i = 1$, the number of expanded paths reaches $2L$ for the main tree and look-ahead tree. Path pruning is then conducted based on the path metrics. The $L$ paths with the largest path metrics are pruned for the main tree, as well as for the corresponding look-ahead paths repreenting $\bv$. %are also pruned and will not be considered for the next stage.  %Therefore, the paths are sorted based on path metrics, and the best $L$ paths are kept for expanding in the next stage. 
    \begin{figure}[ht]
        % \vspace{-5pt}
        \centering
        \includegraphics[width=1\columnwidth]{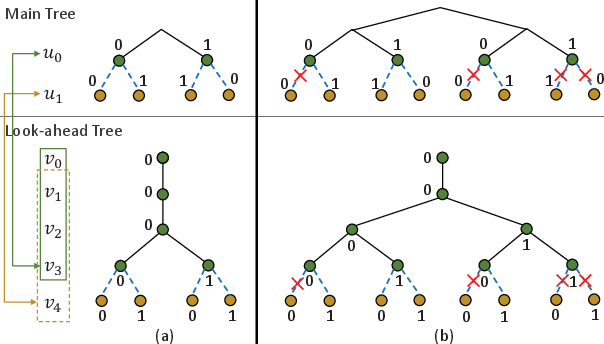}
        \caption{Tree traversal in look-ahead list decoding. }
        \label{fig:RPAC_LASCL_example}
        % \vspace{-5pt}
    \end{figure}   
\end{example}

Given the fundamental idea behind the look-ahead list decoding, we review the computation of path metric, which is identical to the conventional list decoding \cite{rowshan-pac1}. To find a path with the highest probability through the tree traversal restricted to $L$ paths, the following  probability should be {\em{maximized}}:
\begin{equation} %\small
\label{eq:pac_metric1}
\begin{multlined}
%P(\hat{\mathbf{x}}=\hat{\mathbf{u}}\mathbf{P}_n=\hat{\mathbf{v}}\mathbf{G}\mathbf{P}_n|\mathbf{y})=
P(\hat{\mathbf{u}}|\mathbf{y})=\prod_{i=0}^{N-1} P(\hat{u}_i|\hat{u}_0^{i-1},y_0^{N-1}).
\end{multlined}
\end{equation}
As the addition operation is preferable in practice, we define the {\em path metric} for the $l$-th sequence/path $\hat{u}_0^{i-1}$ based on the logarithm of (\ref{eq:pac_metric1}) as %Consider now a partial sequence   $\hat{u}_0^{i-1}=[\hat{u}_0\;\; \hat{u}_1\;\;\ldots\;\; \hat{u}_{i-1}]$ as the the output of the reverse convolutional transform. This sequence determines a path, or a sequence of states, through the trellis for the code. 
$M_{i-1}(l)=-\sum_{j=0}^{i-1}\log P(\hat{u_j}|\hat{u}_0^{j-1},y_0^{N-1})$, which converts maximization in \eqref{eq:pac_metric1} to minimization problem. %We seek to minimize the path metric for the entire codeword (up to bit $i=N-1$) to maximize  (). 
%Since $\log P(\hat{u_i}|\hat{u}_0^{i-1},y_0^{N-1})<0$, this can also save one bit per metric value (i.e., the sign bit) in the storage.
By extending the sequence $\hat{u}_0^{i}$, or equivalently $\hat{v}_0^{i}$, by one bit, the path metric for this longer sequence becomes
\begin{eqnarray} %\small
M_i(l)&=&-\sum_{j=0}^{i}\log P(\hat{u}_j|\hat{u}_0^{j-1},y_0^{N-1})
\label{eq:pac_metric2}\\
&=&M_{i-1}(l)+\mu_i \label{eq:pac_metric3},
\end{eqnarray}
where the {\em branch metric} $\mu_i$ of the extended $l$-th path is: %(see (19) in \cite{rowshan-pac1}):
%\begin{equation} %\small
%\label{eq:pac_metric3}
%\begin{multlined}
%M_t(s)=M_{t-1}(s^\prime)+\mu_t(s^\prime,s)
%\end{multlined}
%\end{equation}
%The path metric along a path to state $s$ at time $t$ is obtained by adding the path metric to the state $s^\prime$ at time $t - 1$ to the branch metric for an input that moves the encoder from state $s^\prime$ to state $s$. If there is no such input, i.e., $s^\prime$ and $s$ are not connected on the trellis, then the branch metric is considered $\infty$.
%To simplify the arithmetic operation, we can define $\mu_i$ based on the decision LLR $\lambda_0^i(l)$ or simply $\lambda_0^i$. 
%\begin{eqnarray} 
\begin{equation}
\label{eq:pac_metric4}
\begin{multlined}
\mu_i= -\log P(\hat{u}_i|\hat{u}_0^{i-1},y_0^{N-1})%\\=-\log\left(\frac{e^{(1-\hat{u}_i)\lambda_0^i}}{e^{\lambda_0^i}+1}\right) %\nonumber \\ 
%\hspace{-3cm}
%\!\!&=&\!\!
=\log\left(1+e^{-(1-2\hat{u}_i)\lambda_0^i}\right).
\end{multlined}
\end{equation}
The branch metric can be approximated using \cite[Equations (5) and (6)]{rowshan2021list}. 
%\end{eqnarray}
%where the last equality holds only for $\hat{u}_i=$ 0 and\,1. 
% Now, for the value of $\hat{u}_i$ that equals  $h(\lambda_0^i)$,
% \begin{equation}
% \label{eq:sc_hard_decision}
% h(\lambda_0^i) = \begin{dcases*}
%         0 & $\lambda_0^i>0$,\\
%         1 & otherwise.\\
% \end{dcases*}
% \end{equation}
% %$\frac{1}{2}(1-\sgn(\lambda_0^t))$, 
% The term $e^{-(1-2\hat{u}_i)\lambda_0^i}=e^{-|\lambda_0^i|}$ is small and therefore $\log(1+~e^{-|\lambda_0^i|}) \approx 0$. Otherwise, we can approximate $\log(1+e^{|\lambda_0^i|})\approx |\lambda_0^i|$. Thus 
% \begin{equation}
% \label{eq:pm_func} %\footmotesize
% \mu_i=\mu_i(\lambda_0^i,\hat{u}_i)\! \approx \!
% \begin{dcases*}
% 0 & if $\hat{u}_i = h(\lambda_0^i)$,\\ %\frac{1}{2} (1\!-\!\sgn(\lambda^t_0))$ 
% |\lambda^i_0| & otherwise.	\\
% \end{dcases*}
% \end{equation}
Obtaining $M_i(l),l\in[1,2L]$, we retain the $L$ paths with the smallest $M_i(l)$. As observed, the main decoding process relies on $\bu$. 
Expanding the look-ahead sequence $[\hat{v}_{i+1}\;\; \hat{v}_{i+2}\;\; \cdots\;\; \hat{v}_{i+s}]$ to obtain $\hat{u}_i$ does not increase the decoding complexity in the main decoding procedure.
%It turns out that this branch metric is equivalent to the one suggested for the list decoding of polar codes in \cite{yuan,balatsoukas} and PAC codes in \cite{rowshan-pac1}.

\section{Numerical Results and Discussions} \label{sec:results}
%In this section, we consider several example codes and numerically evaluate the proposed scheme in terms of error coefficient and block error rate followed by discussions. 
The block error rates (BLER) of the codes (64, 50) and (128, 110) are shown in Figs. \ref{fig:bler_(64,50)} and \ref{fig:bler_(128,110)}, constructed using approximate density evolution method \cite{Urbanke_polar_2010}. %, Trifonov_efficient_2012}.  
We adopt the proposed LA-SCL decoder and the 3rd-order ordered statistic decoder (OSD \cite{Fossorier_Soft_OSD_1995}, denoted by OSD(3), as near-ML decoder to decode the RPAC codes. The error correction performance of RPAC codes is compared with that of polar and PAC codes with SCL decoders and CRC-polar codes with SCL decoders.
Let SCL($L$) and LA-SCL($L$) denote the SCL and LA-SCL decoders with list size $L$. We use PAC($s+1$) to denote conventional PAC codes precoded by polynomials with lengths $s+1$. The same polynomial mentioned in Example \ref{fig:example_rev} is used for RPAC(10) and $\bp = 
    % [1\;1\;0\;1\;1\;0\;1]$ 
    [1\;1\;0\;1\;1\;0\;1]$ is used for RPAC(7). %(except for the constraint length $s+1=4$ where $\bp_4 = [1\;0\;1\;1]$).
The adopted CRC-polar codes use 11 CRC bits with the generator polynomial $g(x) = x^{11} +x^{10} +x^{9} +x^{5} +1$. Table \ref{tb:Admin} gives the minimum distance with the corresponding error coefficient for the underlying codes. %Note that concatenating these high-rate short codes with a short CRC can result in significant performance degradation due to a large rate loss. 

%A table to show the error coefficient of the codes before and after precoding, and after code modification.

% \begin{table}[ht] 
% % \vspace{-5pt}
% \setlength{\tabcolsep}{0.6em} % for the horizontal padding
% \renewcommand{\arraystretch}{1.2} %<- row spacing

% \caption{Minimum weight $\wm$ and the corresponding error coefficient $\Awm$ of polar codes ($^*$ refers to SR-PAC).}\label{tb:Admin}
% \centering
% \begin{tabular}{|l|l|l|l|l|}%llllllll} 
% \cline{1-5}
% {code} & {(64, 50)} & {(64, 32)} & {(64, 14)} & {(128, 110)} \\ 
% \cline{1-5}
% $\Awm$  & $A_{4}$           & $A_{8}$    & $A_{16}$   & $A_{4}$                \\  
% \cline{1-5}
% Polar                                        & 944   & 664   & 172     & 4099 \\

% \cline{1-5}
% SR/R-PAC(4)                                  & 435   & 339   & 220$^*$   & 1621 \\ 

% \cline{1-5}
% SR/R-PAC(7)                                  & 98   & 377$^*$  & 137$^*$    & 240  \\

% \cline{1-5}

% SR/R-PAC(10)                                 
% & 70$^*$     & 107$^*$       & 73$^*$                & 99$^*$                        \\
% \cline{1-5}
% \end{tabular}
% %\vspace{-5pt}
% \end{table}

% % table: 64,14, with Wmin
\begin{table}[ht] 
\vspace{2pt}
\setlength{\tabcolsep}{0.6em} % for the horizontal padding
\renewcommand{\arraystretch}{1.2} %<- row spacing
\caption{Minimum weight $\wm$ and the corresponding error coefficient $\Awm$ of polar codes%($^*$ refers to SR-PAC)
.}
\label{tb:Admin}
\centering
\begin{tabular}{|l|l|l|l|l|}
\cline{1-5}
\multicolumn{1}{|c|}{\multirow{2}{*}{code}} & \multicolumn{2}{c|}{(64, 50)} %& \multicolumn{2}{c|}{(64, 14)} 
& \multicolumn{2}{c|}{(128, 110)} \\

\cline{2-5}
\multicolumn{1}{|c|}{}       & $\wm$ & $\Awm$                  %& $d_{min}$ & $\Adm$                  
& $\wm$ & $\Awm$           \\

\cline{1-5}
Polar                                          
& 4    & 944      %& 16   & 172     
& 4    & 4448     \\

\cline{1-5}
CRC-Polar                                   
& 2    & 2     %& 16   & 73        
& 4    & - \\

\cline{1-5}
PAC(10)                                          
& 4    & 944      %& 16   & 140 
& 4    &  4320    \\

%\cline{1-7}
%FC-PAC(4)                                     
%& 4    & 435      & 16   & 220     & 4   & 1621      \\

% \cline{1-5}
% FC-PAC(7)                            that the     
% & 4    & 98      %& 16   & 137       
% & 4    & 240  \\

\cline{1-5}
RPAC(10)                                   
& 4    & 70     %& 16   & 73        
& 4    & 99  \\

\cline{1-5}

\end{tabular}
 \vspace{-1pt}
\end{table}

% simulation settings

\begin{figure}[ht]
    % \vspace{-5pt}
    \centering
    \includegraphics[width=0.9\columnwidth]{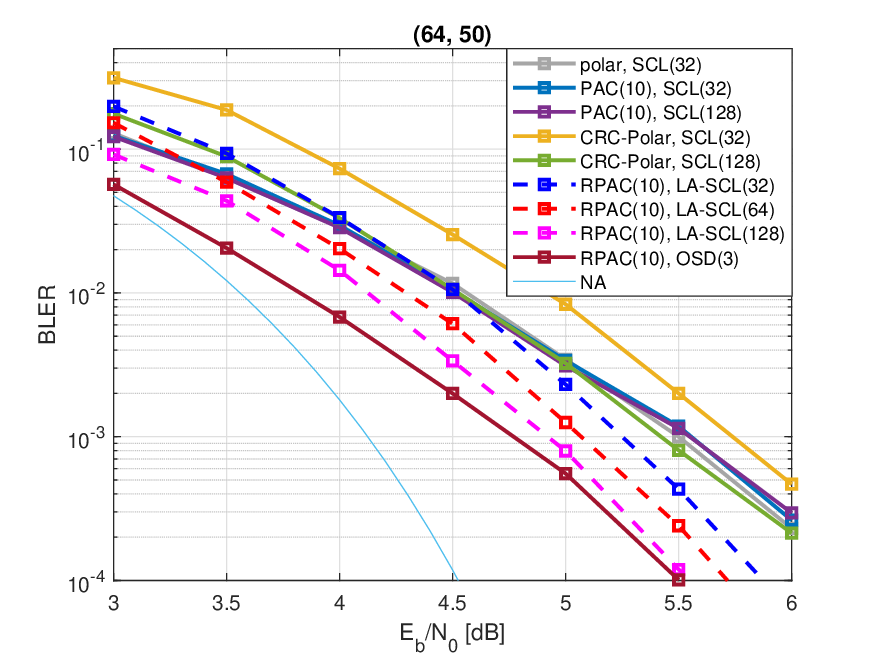}
    \caption{Performance comparison of (64,50) codes.}
    \label{fig:bler_(64,50)}
    % \vspace{-5pt}
\end{figure}

In Fig. \ref{fig:bler_(64,50)}, SCL decoding of PAC code (64,50) with list size $L = 32,128$ and SCL decoding of polar code (64,50) with $L = 32$ show identical error correction performance. 
%As illustrated in Table \ref{tb:Admin}, the precoding in PAC code (64,50) does not change the error coefficient of polar code (64,50). In this case, increasing the list size in the SCL decoding cannot improve the performance of the high-rate PAC code. 
When adopting the CRC-polar code, the minimum distance of the code reduces due to the employment of CRC bits and occupation of more bit-channels, resulting in the degradation of error performance under SCL decoding, as shown in Fig. \ref{fig:bler_(64,50)}. 
With the proposed look-ahead SCL (LA-SCL) decoding, RPAC code (64,50) outperforms the counterpart PAC code and CRC-polar code. From Table \ref{tb:Admin}, it can be observed that %$\Awm$ with reverse pre-transformation, 
more than 92.5$\%$ of the minimum weight codewords of RPAC(10) relative to PAC code are eliminated, leading to  
%Relating the error coefficient to the BLER curve, %for RPAC code, about 0.5 dB power gain can be achieved under LA-SCL(32) decoder compared to CRC-polar code under SCL(32) decoder. By increasing the list size, 
the improvement of up to 0.6 dB % is achievable 
for LA-SCL decoder (for RPAC code) and SCL decoder (for CRC-polar code) when $L = 128$. In high SNR regimes, the performance of RPAC code under SCL(128) decoder approaches that of under OSD decoder and the approximation for BLER under AWGN channel given by normal approximation (NA) \cite{polyanskiy}.

\begin{figure}[ht]
    % \vspace{-5pt}
    \centering
    \includegraphics[width=0.9\columnwidth]{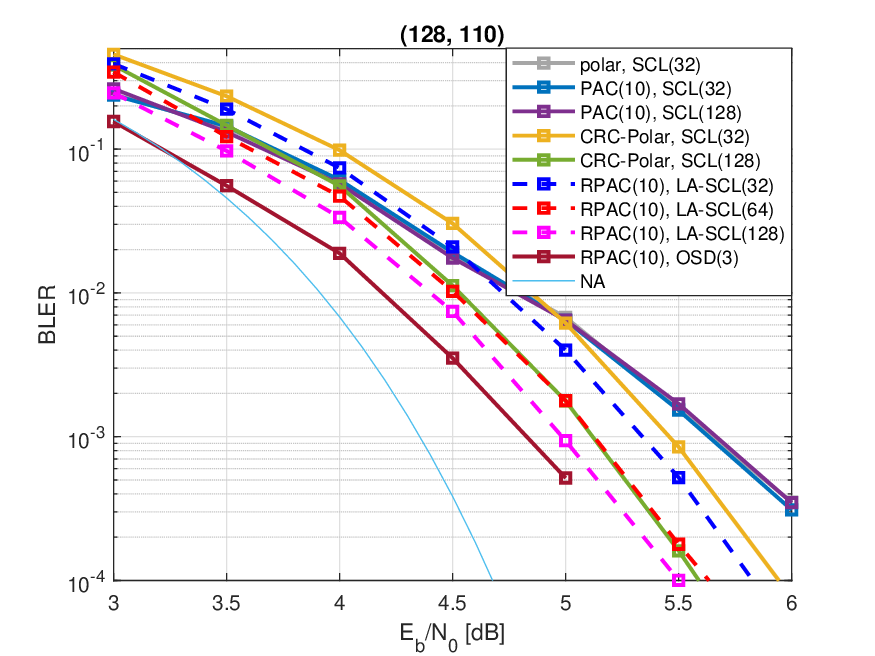}
    \caption{Performance comparison of (128,110) codes.}
    \label{fig:bler_(128,110)}
    % \vspace{-5pt}
\end{figure}

%Similar conclusions can be drawn for longer codes at high rates, e.g., (128,110). As illustrated in Table \ref{tb:Admin} and Fig. \ref{fig:bler_(128,110)}, the considerable decrease in $\Awm$ for RPAC(10) enables the code to outperform its counterpart, PAC code. For CRC-Polar codes, the incorporated CRC polynomial introduces a rate loss in high-rate codes and involves very low-reliability bit-channels, thereby preventing the SCL decoder from significantly enhancing the error correction capabilities of polar codes. CRC-polar code with SCL decoder gives a sharp slope as the SNR increases, leading to a small gain between the proposed LA-SCL decoder for RPAC code. As a result, with $L=32$, about 0.1 dB improvement can be achieved when employing the LA-SCL decoder for the RPAC code. The LA-SCL(64) decoder for RPAC code demonstrates the same performance as the performance of SCL(128) decoder for the CRC-polar code and up to 0.2 dB power gain improvement can be achieved when the list size of LA-SCL decoder is increased to 128, bringing the curve closer to the performance of OSD for RPAC code.

Similar trends are observed for longer codes at high rates, e.g., (128,110). Table \ref{tb:Admin} and Fig. \ref{fig:bler_(128,110)} demonstrate a significant decrease in $\Awm$ for RPAC(10), enabling the code to outperform the PAC code counterpart. The LA-SCL decoder for the RPAC code shows improvements over the CRC-polar code, with gains of up to 0.1 dB achievable with $L=32$. As the list size of the LA-SCL decoder increases to 64 and 128, the performance approaches that of OSD for the RPAC code, with gains of up to 0.2 dB. %possible with a list size of 128, bringing the curve closer to OSD performance.

%As is shown in Fig. \ref{}, 

% \begin{figure}[ht]
%     % \vspace{-5pt}
%     \centering
%     \includegraphics[width=0.9\columnwidth]{64_14_sum_results_globecom.eps}
%     \caption{Performance comparison of (64,14) codes.}
%     \label{fig:bler_(64,14)}
%     % \vspace{-5pt}
% \end{figure}
%For the low-rate code (64,14), limited by a relatively small reduction in $\Awm$, the performance gain is not as significant as high-rate codes. As shown in Fig. \ref{fig:bler_(64,14)}, the low-rate CRC-polar code under the SCL decoder approaches the performance of the FC-PAC code as the SNR increases. Compared to low-rate codes under SCL and sphere decoding schemes, FC-PAC code can provide about 0.2 dB power gain at relatively lower and medium SNR regimes. Our observations show that the potential for error coefficient improvement increases with an increase in code rate. The reason is that low-rate codes usually have a small error coefficient due to a limited number of cosets $\Ci,i\in\B$ in these codes. Hence, a significant reduction in the error coefficient is hard to achieve. 

%Meanwhile, the benefit of following one decoding path, SD requires less memory and resources compared to SCLD(8).
%Discussion on the performance comparison based on the table below: 

%%%%%%%%%%%%%%%%%%%%%%%%%%%%%%%%%%%%%%%%%%%%%%%%%%%%%%%%%%%%%%%%%%%%%
\section{CONCLUSION} %AND FUTURE DIRECTIONS}
In this paper, we introduce the reverse PAC coding and its associated look-ahead list decoding. 
%This precoding scheme for polar codes can overcome the limitations of PAC coding and break the lower bound for the number of minimum weight codewords \cite[Lemma 3]{rowshan2023minimum}.
%This goal is achieved by involving constructive frozen rows in row combinations that results in the formation of codewords with weights greater than $\wm$. 
The RPAC codes have a remarkably smaller number of minimum-weight codewords compared to polar codes and conventional PAC codes for high-rate codes. 
%The look-ahead list decoding is proposed for decoding RPAC codes that with a similar complexity as conventional list decoding can outperform PAC codes and CRC-Polar codes at high rates. 
The proposed LA-SCL decoder overcomes the challenge posed by the opposite directions of demapping and decoding in RPAC codes. With a similar complexity as conventional list decoding, the LA-SCL decoder for RPAC codes outperforms CRC-Polar codes and PAC codes under conventional list decoding for high-rate short codes.

%Knowing that employing CRC with high-rate short codes can introduce a large rate loss and result in significant performance degradation. 
%This reduced error coefficient results in the considerable improvement of the BLER under sphere decoding, in particular for high-rate codes. 
%The future direction could be further exploitation of the frozen cosets and reshuffling the other cosets to improve the weight distribution. %For high rate codes, the proposed schemes allow the codes to approach the union bound.
%Since this work is focused on improving the code performance by precoding, the techniques to reduce the complexity and design code constructions for each precoding scheme can be considered as the direction of future works. 
%The techniques to design the row combinations with frozen cosets of polar codes to further reduce the error coefficients 
%reducing the error coefficients of the codes by precoding
%can be considered as the direction of future works. 
%Also, a study on different  can be performed as future work. %of this scheme. 

\addtolength{\textheight}{-12cm}   % This command serves to balance the column lengths
\end{document}